\documentclass[fleqn,usenatbib]{mnras}

\usepackage{newtxtext,newtxmath}

\usepackage[T1]{fontenc}

\DeclareRobustCommand{\VAN}[3]{#2}
\let\VANthebibliography\thebibliography
\def\thebibliography{\DeclareRobustCommand{\VAN}[3]{##3}\VANthebibliography}

\usepackage{graphicx}	
\usepackage{amsmath}	
\usepackage[modulo]{lineno}

\defcitealias{liu2019formation}{L19}

\usepackage{xspace}
\newcommand{\flash}{{\sc flash}\xspace}
\newcommand{\swift}{{\sc Swift}\xspace}
\newcommand{\woma}{{\sc WoMa}\xspace}
\newcommand{\seagen}{{\sc SEAGen}\xspace}

\title[No dilute core in simulations]{No dilute core produced in simulations of giant impacts on to Jupiter}

\author[T. D. Sandnes et al.]{
T.~D.~Sandnes,$^{1,}$\thanks{E-mail: thomas.d.sandnes@durham.ac.uk}
V.~R.~Eke,$^{1}$
J.~A.~Kegerreis,$^{2,3,4}$
R.~J.~Massey,$^{1}$
and L.~F.~A.~Teodoro,$^{5,6}$
\\
$^{1}$Institute for Computational Cosmology, Department of Physics, Durham University, South Road, Durham, DH1 3LE, UK\\
$^{2}$Department of Earth Science and Engineering, Imperial College London, London, SW7 2BP, UK\\
$^{3}$SETI Institute, 339 Bernardo Avenue, Suite 200, Mountain View, CA 94043, USA\\
$^{4}$NASA Ames Research Center, MS 245-3, Moffett Field, CA 94035, USA\\
$^{5}$Faculty of Mathematics and Natural Sciences, University of Oslo,
 Sem S{\ae}lands vei 24, 0371 Oslo, Norway\\
$^{6}$School of Physics and Astronomy, University of Glasgow, G12 8QQ, Scotland, UK
}

\date{Accepted XXX. Received YYY; in original form ZZZ}

\pubyear{2024}

\begin{document}
\label{firstpage}
\pagerange{\pageref{firstpage}--\pageref{lastpage}}
\maketitle

\begin{abstract}
A giant impact has been proposed as a possible formation mechanism for Jupiter's dilute core -- the planet's inferred internal structure in which the transition between its core of heavy elements and its predominantly hydrogen--helium envelope is gradual rather than a discrete interface. A past simulation suggested that a head-on impact of a $10~M_\oplus$ planet into an almost fully formed, differentiated Jupiter could lead to a post-impact planet with a smooth compositional gradient and a central heavy-element fraction as low as $Z\approx0.5$. Here, we present simulations of giant impacts on to Jupiter using improved numerical methods to reassess the feasibility of this scenario. We use the REMIX smoothed particle hydrodynamics (SPH) formulation, which has been newly developed to improve the treatment of mixing in SPH simulations. We note that, as in previous works, chemical mixing is not included in these models and that incorporating such processes at sub-particle scales could improve numerical convergence. We perform giant impact simulations with varying speeds, angles, pre-impact planet structures, and equations of state. In all of our simulations, heavy elements re-settle over short time-scales to form a differentiated core, even in cases where the core is initially disrupted into a transiently mixed state. A dilute core is not produced in any of our simulations. These results, combined with recent observations that indicate Saturn also has a dilute core, suggest that such structures are produced as part of the extended formation and evolution of giant planets, rather than through extreme, low-likelihood giant impacts.
\end{abstract}





\section{Introduction}\label{Jupiter_sec:introduction}

Measurements of Jupiter's gravitational moments by the Juno spacecraft have led to models of the planet's interior that suggest the existence of a dilute core: an extended compositional gradient between Jupiter's central core of heavy elements and its hydrogen--helium envelope \citep{wahl2017comparing, nettelmann2017low, vazan2018jupiter, debras2019new, militzer2022juno, miguel2022jupiter, howard2023jupiter, militzer2024}. This is inconsistent with traditional giant planet formation models that predict a differentiated internal structure \citep{muller2020challenge}. With ring seismology suggesting that Saturn also has a dilute core \citep{mankovich2021diffuse}, understanding the processes that govern the formation of such compositional gradients would provide key insights into the evolution of giant planets and planetary systems.

Several mechanisms have been proposed to explain Jupiter's dilute core \citep{helled2022revelations}. An extended planetesimal-dominated accretion phase could lead to the dilute core being in place prior to runaway gas accretion \citep{venturini2020jupiter,Stevenson+2022}. Alternatively, convective processes could gradually erode a differentiated core until it reaches a mixed state \citep{moll2017double}. \citet{liu2019formation} (hereafter \citetalias{liu2019formation}) proposed a giant impact as an alternative mechanism.

The head-on impact simulation of \citetalias{liu2019formation} presented the disruption of a differentiated core by a $10~M_\oplus$ impactor into a well-mixed, diluted state with a heavy-element fraction in the centre of the planet of ${Z<0.5}$. This extreme giant impact would deliver approximately half of the planet's heavy elements in a single event. The simulations of \citetalias{liu2019formation} with a larger impact parameter or a smaller impactor mass did not produce a dilute core. The hydrodynamic simulations of \citetalias{liu2019formation} were carried out using the adaptive mesh code \flash \citep{fryxell2000flash}. By separately modelling the subsequent thermodynamic evolution, \citetalias{liu2019formation} found that this compositional gradient could persist for Gyr time-scales until the present day. However, overmixing in regions of large bulk motion relative to the stationary grid points is a typical shortcoming of Eulerian methods \citep{springel2010pur, robertson2010computational}. This spurious diffusion arises from the advection terms necessary in this non-Lagrangian method. Additionally, the accuracy of the treatment of self-gravity is sensitive to choices made in the multipole approximation of the gravitational potential \citep{couch2013improved}, and \citetalias{liu2019formation} used idealised equations of state (EoS) that do not capture the complexities of metallic hydrogen within Jupiter's deep interior \citep{chabrier2019new}. Therefore, further investigation is warranted of the impact scenario using fundamentally different modelling approaches, to assess the potential sensitivity of dilute core production to the specifics of the numerical methods employed. 

Lagrangian hydrodynamic methods, where Galilean invariance is maintained, do not experience the artificial mixing from advection through grid points that is observed in methods that use a stationary mesh, since interpolation points move with the fluid velocity. In particular, smoothed particle hydrodynamics (SPH) is widely used for simulations of giant impacts since it: inherently tracks the evolution of fluid element trajectories and thermodynamics; is able to deal with vacuum regions and evolving free surfaces efficiently; offers geometry-independent adaptive resolution; and couples elegantly with gravity solvers \citep{lucy1977numerical, gingold1977smoothed}. In traditional SPH (tSPH) formulations, however, mixing at density discontinuities is typically suppressed by spurious surface tension-like effects \citep{agertz2007fundamental}. These artificial effects are considerable -- and challenging to remedy -- at boundaries between dissimilar, stiff materials, for which more significantly erroneous estimates of fluid pressure suppress mixing more strongly  \citep{ruiz2021effect}. Therefore, a more advanced SPH construction that addresses these known sources of error is needed to utilise the benefits of the SPH formulation to reliably investigate dilute core formation in giant impact simulations like these where material mixing is the key physical mechanism of interest. 
 
REMIX is an advanced SPH scheme designed to directly address the sources of numerical error that suppress mixing in SPH simulations \citep{sandnes2025remix}. The REMIX scheme incorporates a range of novel and recently developed improvements to tSPH formulations, and its construction is generalised to address sources of error independent of material type or EoS. It demonstrates significant improvements in the treatment of both mixing and instability growth, including in simulations with materials and conditions representative of those in giant impact simulations \citep{sandnes2025remix}. REMIX is integrated into the open-source, state-of-the-art \swift code\footnote{\swift \citep{schaller2024swift} is publicly available at \url{www.swiftsim.com}.}, whose computational efficiency enables simulations of planetary giant impacts to be performed at high resolutions \citep[e.g.,][]{kegerreis2022immediate}.

Here, we use REMIX SPH to investigate whether Jupiter's dilute core could be formed by a giant impact. First, in \S\ref{Jupiter_sec:methods} we describe the methods used to perform simulations and to construct initial conditions. In \S\ref{Jupiter_sec:instabilities}, we test REMIX in fluid instability simulations under conditions representative of Jupiter's deep interior. Then, in \S\ref{Jupiter_sec:giant_impacts} we use REMIX to model giant impacts on to Jupiter. We carry out simulations of: head-on impacts (\S\ref{Jupiter_subsec:fiducial}); isolated planets with a pre-constructed dilute core to assess the potential stability of dilute-core structures in our simulations (\S\ref{Jupiter_subsec:predilutecore}); impacts at a range of impact speeds and angles (\S\ref{Jupiter_subsec:speed_angle}); and impacts with pre-impact planet structures and EoS set up to closely follow and compare with those of \citetalias{liu2019formation} (\S\ref{Jupiter_subsec:L19_impact}). We discuss our results in \S\ref{Jupiter_sec:discussion} and summarise our findings in \S\ref{Jupiter_sec:conclusions}.

\section{Methods}\label{Jupiter_sec:methods}

\subsection{REMIX smoothed particle hydrodynamics}\label{Jupiter_subsec:remix}

REMIX is an SPH formulation designed to address key sources of error that suppress mixing and instability growth in tSPH simulations, particularly at density discontinuities. By adopting a generalised, material-independent approach, REMIX is able not only to improve the treatment of contact discontinuities within a single material but also to handle well the more challenging case of interfaces between dissimilar, stiff materials. Like tSPH, REMIX inherently conserves mass, energy, and momentum; is constructed from a basis of thermodynamic consistency; and is fully Lagrangian, ensuring Galilean invariance. REMIX has been extensively tested with standard hydrodynamics and giant impact-relevant test scenarios, with full details presented in \citet{sandnes2025remix}. Here, we summarise the primary features of REMIX and set up some additional test scenarios tailored directly to a Jupiter core-mixing context.
  
In tSPH formulations used for applications in astrophysics, the fluid density at the positions of particles is estimated by kernel interpolation using an extended, Gaussian-like kernel function \citep{price2012smoothed}. The standard SPH density estimate will smooth the density field on kernel length-scales. In regions where the density varies smoothly, this will be a minor effect. However, in sharply varying regions, and in particular at discontinuities in the underlying field, the reconstructed density field will inevitably be smoothed. 

The effect of this kernel smoothing can be clearly seen at interfaces between different material layers in the pre-impact planets used for our giant impact simulations. Pre-impact planetary equilibrium profiles and the corresponding SPH particle placements are calculated using the publicly available \woma\footnote{The \woma code \citep{ruiz2021effect} for producing spherical and spinning planetary profiles and initial conditions is publicly available with documentation and examples at \href{https://github.com/srbonilla/WoMa}{github.com/srbonilla/WoMa}, and the python module \texttt{woma} can be installed directly with \href{https://pypi.org/project/woma/}{pip}.} and \seagen\footnote{\seagen \citep{kegerreis2019planetary} is publicly available at \href{https://github.com/jkeger/seagen}{github.com/jkeger/seagen}, or as part of \woma.} codes. Prior to impact simulations, additional adiabatic ``settling'' simulations are performed to allow  particles to rearrange themselves towards an equilibrium configuration. In these simulations, particle entropies are fixed to their initial value to enforce adiabatic evolution. Settling simulations are carried out separately for each planet and are run for a simulation time of 5000~s. REMIX reduces the errors that traditionally make calculations of particle accelerations sensitive to the local particle configuration. Therefore, the amount of particle motion in settling simulations is reduced.

\begin{figure}
	\centering
{\includegraphics[width=0.48\textwidth, trim={2.5mm 0mm 2.5mm 0mm}, clip]{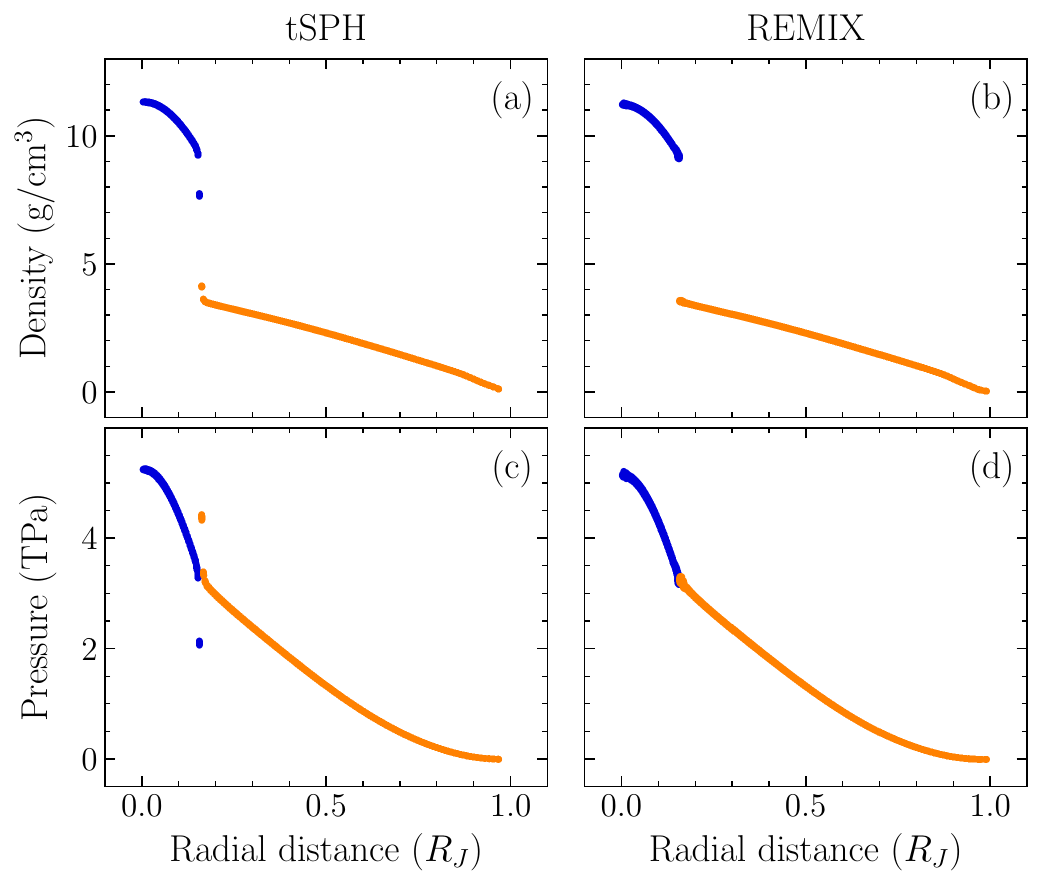}
}\hfill
    \vspace{-2em}
	\caption{Radial profiles of density (a, b) and pressure (c, d) for a two-layer proto-Jupiter settling simulation at time $t = 5000$~s. Columns show profiles from simulations using a tSPH formulation (a, c) and REMIX SPH (b, d). Individual particles are coloured by material type: blue for the central core of ice and orange for the hydrogen--helium envelope.}
	\label{Jupiter_fig:profiles}
\end{figure}

The radial density and pressure profiles of a proto-Jupiter planet, to be used in our impact simulations, is shown in Fig.~\ref{Jupiter_fig:profiles} from settling simulations using tSPH and REMIX. With tSPH the density field is smoothed by kernel interpolation, leading to diverging pressures at the core--envelope boundary that act as an artificial barrier to mixing across the interface. With REMIX, the density discontinuity stays sharp and the pressure remains continuous across the material boundary. To address kernel smoothing error, REMIX uses a differential form of the density estimate by which particle densities are evolved with time rather than recalculated from the instantaneous distribution of particle masses. All density discontinuities, including those at material boundaries and free surfaces, are not erroneously smoothed.

The smoothing error introduced by using an extended kernel function combines with the error introduced by the discretisation of the underlying fluid into a finite set of particles \citep{price2012smoothed, spreng2020advanced}. To deal with discretisation error and mitigate the accumulation of error over time in evolved densities and internal energies, REMIX uses: linear-order reproducing kernels \citep{frontiere2017crksph} that adapt to treat free surfaces as vacuum boundaries; a choice of free functions in the SPH equations of motion that limits discretisation error \citep{read2010resolving}; a kernel normalising term in the density evolution calculations; advanced formulations of artificial viscosity as well as artificial diffusion of internal energy and density between materials of the same type.

All simulations presented here use particles of equal mass across the simulation, a case that was specifically considered in the validation of REMIX \citep{sandnes2025remix}, and we employ the Wendland $C^2$ kernel with $\eta=1.487$ to construct linear-order reproducing kernels \citep{wendland1995piecewise, dehnen2012improving}. Particles have approximately 100 neighbours within their kernel. REMIX has been developed with computational efficiency in mind and therefore, as demonstrated by the simulations presented here, can be used in simulations at state-of-the-art resolutions for giant impact simulations.

\subsection{Equations of state}\label{Jupiter_subsec:EoS}

The EoS characterise the thermodynamic behaviour of a material. In the SPH simulations presented here, the EoS are used to calculate pressures and sound speeds from densities and internal energies. These quantities are then used both to directly evolve the simulated fluid and in calculations of time-step durations. 

Models of Jupiter's internal structure are sensitive to uncertainties in the hydrogen--helium EoS used to calculate the planet's envelope profiles \citep{miguel2016jupiter, mazevet2022benchmarking, howard2023jupiter}, with much work ongoing to create EoS that accurately reproduce the behaviour of hydrogen--helium at the extreme densities and pressures in the interiors of
giant planets \citep{saumon1995equation, militzer2013ab, chabrier2019new}. For simulations of giant impacts on to Jupiter and hydrodynamic tests using Jupiter-like materials, we use the \citet{chabrier2021new} hydrogen--helium EoS (hereafter CD21 H--He), with a helium mass fraction of $Y = 0.245$ \citep{chabrier2019new}. For simulations of impacts on to Jupiter aiming to reproduce directly the initial conditions of \citetalias{liu2019formation}, we use an ideal gas with adiabatic index $\gamma = 2$.

For heavy elements, we use the AQUA EoS \citep{haldemann2020aqua} to represent ice and the ANEOS forsterite EoS \citep{stewart2020shock} for rocky material. For direct \citetalias{liu2019formation} comparison simulations, we use Tillotson ice and granite \citep{melosh1989impact}.

\subsection{Impact initial conditions}\label{Jupiter_subsec:ICs}

For the majority of our simulations, we use a differentiated two-layer proto-Jupiter with a heavy-element core of ice and a H--He envelope, and a single-layer ice impactor. For simulations set up to most closely match the initial conditions of the simulations of \citetalias{liu2019formation}, we use three-layer pre-impact planets with layers of rock, ice, and gas for both target and impactor. The choice to focus on impacts between planets with a reduced number of layers is made to further reduce any potential barriers to mixing. In all of our simulations, we follow \citetalias{liu2019formation}'s scenario and use an impactor with a total mass of $10~M_\oplus$ and a proto-Jupiter of total mass $308~M_\oplus$, with core mass of ${\sim}10~M_\oplus$, where $M_\oplus=5.972\times10^{24}$~kg. The total mass of the system is therefore the present-day mass of Jupiter, $M_J = 1.898\times10^{27}$~kg. Some simulations with three-layer planets have a slightly more massive core of $11.6~M_\oplus$ to give equilibrium profiles that more closely match those of \citetalias{liu2019formation}, with different EoS, although we find that changes to the initial profiles do not significantly affect the evolution of the impact. H--He layers are chosen to be adiabatic with surface temperatures (defined by where $P=1$~bar) of 165~K for proto-Jupiters and 500~K for impactors. For two-layer proto-Jupiters and single-layer impactors, the ice layer is also chosen to be adiabatic with an impactor surface temperature of 200~K. For three-layer planets, the temperature--density relation of heavy-element layers is chosen somewhat arbitrarily, to attempt to match the radii of material interfaces of the simulations of \citetalias{liu2019formation}: all are isothermal except for the impactor ice layer which has $T\sim\sqrt{\rho}$.

All impact simulations are performed in 3D. They are set up 1~h prior to impact as detailed in \citet[][appendix B.2]{Kegerreis+2025}, defined as the planets' individual centres of mass reaching the summed distance of their initial radii, such that the shapes of the planets are allowed to realistically distort under tidal forces. At Jupiter's orbital distance from the Sun, we expect the peculiar velocity of the impactor to be small compared with the mutual escape speed, $v_{\rm esc} = 54$~km~s$^{-1}$, and therefore we simulate most impacts with an impact velocity of $v = v_{\rm esc}$. Note that \citetalias{liu2019formation} simulate impacts with $v = 46$~km~s$^{-1}$ at the point of impact. Our impact parameter space exploration includes speeds as low as $v = 40.5$~km~s$^{-1}$ to test the potential implications of this choice.

We carry out a suite of simulations to systematically probe the effect of impact speed ($v$ = 0.75, 1.0, 1.5~$v_{\rm esc}$); impact angle (with impact parameter $b$ = 0.0, 0.2, 0.4, 0.6); and numerical resolution (with particle number $N$ = $10^5$--$10^{8}$ in logarithmic steps of $10^{0.5}$). These simulations are based on our fiducial simulation that uses planets with a reduced number of layers, is head-on, is at the mutual escape velocity, and has resolution $10^7$. We also carry out simulations to replicate the impact of \citetalias{liu2019formation} even more closely, with three-layer bodies using the EoS used in their simulations, as well as with the more sophisticated EoS detailed above.

\subsection{Measures of material mixing}\label{Jupiter_subsec:measure_mixing}

Parametrising material mixing will enable us to quantitatively describe the degree to which core material may be diluted throughout the impact. We measure the state of mixing in our impact simulations using two parameters: the local heavy-element mass fraction, $\bar{Z}$, and the total mass of mixed material across the simulation, $M_{\rm mix}$. These quantities describe the local and global state of material mixing respectively. In our simulations, mixing is treated at the particle scale and not below. Therefore, the material of each particle remains fixed for the duration of the simulation. To estimate mixing, we therefore calculate these quantities as weighted estimates based on the localised distributions of particle material types.

We use kernel interpolation to estimate the local heavy-element mass fraction. The quantity $\bar{Z}$ is calculated based on weighted contributions from nearby SPH particles. This parameter describes the fraction of local mass that is represented by heavy-element SPH particles, such that $\bar{Z} = 0$ in regions where no local particles are heavy elements and $\bar{Z} = 1$ where they all are. We estimate $\bar{Z}$ at the positions of particles with the standard SPH approach:

\begin{equation}\label{Jupiter_eq:Z}
   \bar{Z}_i \equiv \frac{\sum_{j} \zeta_{j} \, m_j \, W_{ij} V_j}{\sum_{j} m_j \, W_{ij} V_j} \;.
\end{equation}

\noindent
Here subscripts denote quantities either sampled at the position of, or associated with, a particle $i$ or its neighbouring particles $j$. Sums are approximations of integrals over discrete volume elements $V_j = m_j / \rho_j$, where $m_j$ and $\rho_j$ are particle masses and densities. The kernel function $W_{ij} \equiv W(\mathbf{r}_{ij},\, h_i)$ contributes weighting based on the particle separation $\mathbf{r}_{ij} \equiv \mathbf{r}_{i} - \mathbf{r}_{j}$ and is characterised by the smoothing length $h_i$. The parameter $\zeta_j$ takes the value 1 if particle $j$'s material represents heavy elements, and is 0 otherwise. We use the spherically symmetric Wendland $C^2$ kernel function for these calculations \citep{wendland1995piecewise}. We use this rather than the linear-order reproducing kernels used in REMIX, since it provides a more simple and method-independent measure of the mixing.

To estimate the total mass of mixed material in our simulations we first define what constitutes a mixed state. Since each particle retains its material for the duration of the simulation, we determine that a particle with neighbours of different material-types only -- with no neighbours of its own type -- is in a maximally mixed state. For a particle $i$ we estimate the local mass fraction of particle $i$'s own material, similarly to $\bar{Z}$, by

\begin{equation}\label{Jupiter_eq:w}
   \bar{w}_i \equiv \frac{\sum_{j} \kappa_{ij} \, m_j \, W_{ij} V_j}{\sum_{j} m_j \, W_{ij} V_j} \;,
\end{equation}

\noindent
where $\kappa_{ij} = 1$ for particle pairs of the same material and $\kappa_{ij} = 0$ otherwise. We note that, unlike $\bar{Z}$, the value of $\bar{w}$ will never reach 0 because of the contribution of $i$ itself in this calculation. We estimate that the contribution to $m_i$ from materials \textit{different} from that of $i$ to be ${m_{\text{mix}, \, i} \equiv  \left(1 - \bar{w}_i\right) \, m_{i}}$. The total mixed mass in the simulation is then given by

\begin{equation}\label{Jupiter_eq:M_mix}
   M_{\rm mix} \equiv \sum_{i} m_{\text{mix}, \, i} =  \sum_{i} \left(1 - \bar{w}_i\right) \, m_{i}\;,
\end{equation}

\noindent
where we sum over all simulation particles.

We note that both $\bar{Z}$ and $M_{\rm mix}$ will be spatially smoothed on the scale of the smoothing length, since they are calculated by interpolation using an extended kernel. Therefore material near sharp material interfaces will be measured as mixed even if particles of different materials have not crossed the interface.

\section{Fluid instabilities and mixing}\label{Jupiter_sec:instabilities}

Before running the primary impact simulations, we first test REMIX in simulations of Kelvin--Helmholtz instabilities (KHI) and Rayleigh--Taylor instabilities (RTI) with materials and conditions representative of material interfaces in giant impacts on to Jupiter. Although no converged reference solutions exist for these scenarios, demonstrating that REMIX alleviates the purely numerical known issues of tSPH at the material interface will verify that the material-independent improvements of REMIX are effective in this regime, where core-material and metallic hydrogen have been predicted to be miscible \citep{wilson2011solubility, wilson2012rocky}. We carry out fluid instability simulations in 3D and with particles of equal mass across the simulation to validate our hydrodynamic treatment for our impact simulations, as done in \citet{sandnes2025remix} for similar but not Jupiter-specific tests.

\begin{figure*}
	\centering
{\includegraphics[width=\textwidth, trim={2.5mm 0mm 2.5mm 0mm}, clip]{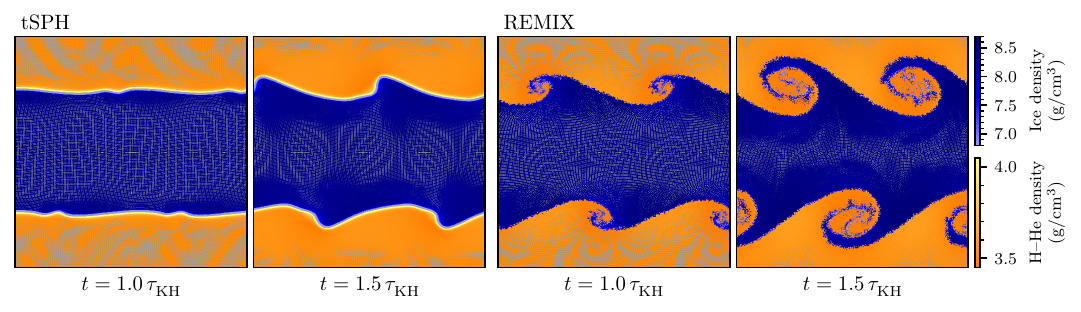}%
}\hfill	
    \vspace{-2em}
	\caption{KHI growth with materials and conditions representative of a pre-impact Jupiter's core--envelope interface. Snapshots show two times from simulations using a tSPH formulation and REMIX. Individual particles are plotted on a grey background and coloured by their material type and density. Particles at all $z$ are plotted, so the grey background is visible in regions that have maintained their grid alignment in $z$ from the initial conditions.}
	\label{Jupiter_fig:KHI}
    \vspace{-1em}
\end{figure*}

\begin{figure*}
	\centering
{\includegraphics[width=\textwidth, trim={2.5mm 0mm 2.5mm 0mm}, clip]{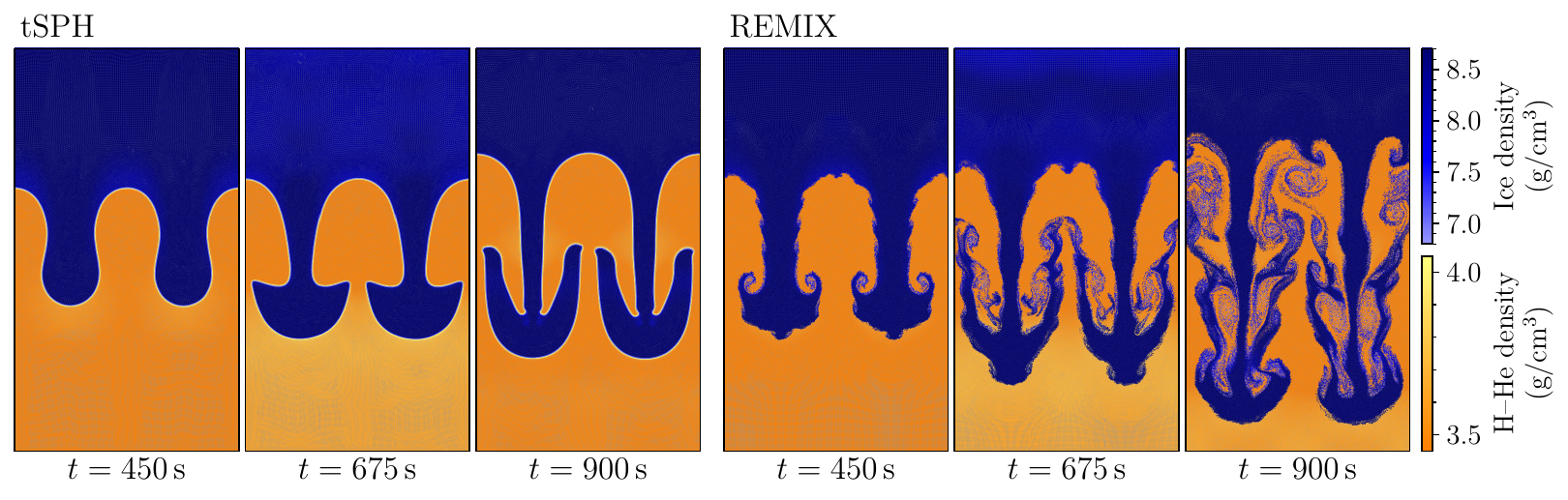}%
}\hfill
    \vspace{-2em}
	\caption{RTI for Jupiter-like materials and conditions, plotted as in Fig.~\ref{Jupiter_fig:KHI}. Snapshots show three times for simulations using a tSPH formulation and REMIX. The regions of fixed boundary particles at the top and bottom of the simulations have been cropped from the figure; their positions and densities do not change.}
	\label{Jupiter_fig:RTI}
    \vspace{-1em}
\end{figure*}

\subsection{Kelvin--Helmholtz instability}\label{Jupiter_subsec:kh}

The KHI arises as perturbations at shearing fluid interfaces grow to form spiralling vortices \citep{chandrasekhar1961hydrodynamic}. We examine the growth of the KHI between layers of ice and H--He
at conditions representative of Jupiter's deep interior. In our simulations these materials are treated as inviscid fluids and so, since the growth of the instability is predominantly inertial, we expect a qualitatively similar evolution to analogous, well-studied ideal gas simulations \citep{price2008modelling, robertson2010computational, mcnally2012well, frontiere2017crksph, rosswog2020lagrangian}. We characterise the growth of a mode of wavelength $\lambda$ by the time-scale

\begin{equation}\label{Jupiter_eq:tau_KH}
   \tau_{\rm KH} = \frac{(\rho_1 + \rho_2) ~\lambda}{\sqrt{\rho_1 \rho_2} \, |v_1 - v_2|} \;,
\end{equation}

\noindent
where $\rho_1$ and $\rho_2$ are the densities in regions separated by the shearing interface and $|v_1 - v_2|$ is their relative speed \citep{price2008modelling}.

Initial conditions are constructed similarly to those of \citet{sandnes2025remix}. H--He particles are initialised in a 3D cubic lattice in a periodic domain with $128 \times 128 \times 18$ particles in the $x, y, z$ directions. The size of the simulation domain in $x$ and $y$ is $1~R_{\rm J}$, where the radius of Jupiter is $R_{\rm J} = 69.9\times 10 ^3$~km, and particle masses are chosen to give a density of $\rho_1 = 3.5$~g~cm$^{-3}$. A region occupying the central half of the domain in $y$ and spanning the full domain in $x$ and $z$ is replaced by a region of higher density ice at $\rho_2 = 8.43$~g~cm$^{-3}$. These densities are chosen to correspond to the densities at the core--envelope interface in the pre-impact proto-Jupiter, as plotted in Fig.~\ref{Jupiter_fig:profiles}.  Since we use particles of equal mass across the simulation, the cubic lattice of ice particles is initialised with a smaller grid-spacing. The particle configurations in both regions are constructed to maintain their grid-spacing across boundaries of the periodic domain and for the two regions to be separated by the mean of the two grid-spacings at both interfaces. The two regions are initialised with relative speeds of $v_1 = -10^{-4}~R_{\rm J}$~s$^{-1}$ and $v_2 = 10^{-4}~R_{\rm J}$~s$^{-1}$. A mode of wavelength $\lambda = 0.5~R_{\rm J}$ and of form  ${v_y = 0.01 |v_1 - v_2| \sin{(2 \pi x / \lambda)}}$ seeds the instability. Initial internal energies are set such that the regions are in pressure equilibrium with $P(\rho, u) = 3.2\times10^{12}$~Pa. We note that the spurious smoothing of the density discontinuity in tSPH means that, unlike with REMIX, simulations with tSPH are not truly initialised in pressure equilibrium.

The evolution of the KHI with these initial conditions, from simulations using tSPH and REMIX, is shown in Fig.~\ref{Jupiter_fig:KHI}. The growth of the instability is clearly and strongly suppressed with tSPH: the characteristic spirals of the KHI do not form and particles are prevented from crossing the density discontinuity by spurious surface tension-like effects. REMIX directly addresses the sources of error that lead to these effects and so allows the instability to grow, and particles of different materials are able to intermix. The instability grows over a similar time-scale, scaled by $\tau_{\rm KH}$, to the analogous KHI simulations with an ideal gas, and also those between Earth-like materials presented in \citet{sandnes2025remix}.

\subsection{Rayleigh--Taylor instability}\label{Jupiter_subsec:rt}

The RTI occurs due to the displacement of a high-density fluid by a low-density fluid \citep{chandrasekhar1961hydrodynamic}. We consider a gravity-driven case in which a region of dense ice sits above a region of H--He, initially in approximate hydrostatic equilibrium other than a small velocity seed perturbation. As in the KHI, spurious surface tension-like effects at the density discontinuity strongly suppress the growth of this instability in tSPH simulations.

Initial conditions are constructed similarly to those of \citet{sandnes2025remix}. Particles are placed in a periodic simulation domain in two cubic lattices. The domain has dimensions of $0.5~R_{\rm J}$, $1~R_{\rm J}$ in the $x$ and $y$ directions, with a thin $3.5 \times 10^{-3}~R_{\rm J}$  domain size in the $z$ dimension. The low density H--He region has $256 \times 256 \times 18$ particles with density $\rho_1 = 3.5$~g~cm$^{-3}$ and occupies the bottom half of the domain. The upper ice region is constructed to satisfy similar grid-spacing constraints as in the KHI simulation, with density  $\rho_2 = 8.43$~g~cm$^{-3}$. Particles in the top and bottom $0.05~R_{\rm J}$ of the domain are fixed in place throughout the course of the simulation. Initial internal energies are set to satisfy hydrostatic equilibrium for a constant gravitational acceleration $g = -31.4$~m~s$^{-2}$, and an interface pressure of $P_0 = 3.2\times10^{12}$~Pa, representative of the gravitational acceleration and pressure at the core--envelope boundary in the proto-Jupiter used for our fiducial giant impact simulations. Particles are initially at rest, other than an initial velocity perturbation that seeds the instability,

\begin{equation}\label{Jupiter_eq:rt_seed}
    v_y(x, y) = \delta_y \left( 1 + \cos\left[8\pi\left(x + 0.25\right) \right] \right) \left(1 + \cos\left[ 5\pi \left(y - 0.5\right) \right] \right)
\end{equation}

\noindent
in the region $0.3~R_{\rm J} < y < 0.7~R_{\rm J}$ and $v_y = 0$ otherwise. We use a perturbation amplitude of $\delta_y = 0.025~R_{\rm J}$~s$^{-1}$.

The evolution of the RTI with these initial conditions is shown in Fig.~\ref{Jupiter_fig:RTI}, for simulations using tSPH and REMIX. In the tSPH simulation, the RTI plumes grow slowly and material is prevented from crossing the interfaces. This is in contrast with the REMIX RTI, where we observe the unimpeded growth of both the primary and secondary instabilities. This leads to mixing across a range of length-scales as particles are not artificially prevented from crossing the interface and instabilities grow to drive turbulent mixing.

The results of these KHI and RTI simulations demonstrate that REMIX does not suppress mixing and fluid instability growth in conditions representative of the giant impact simulations presented in the following section. 

\begin{figure*}
	\centering
{\includegraphics[width=\textwidth, trim={2.5mm 0mm 2.5mm 0mm}, clip]{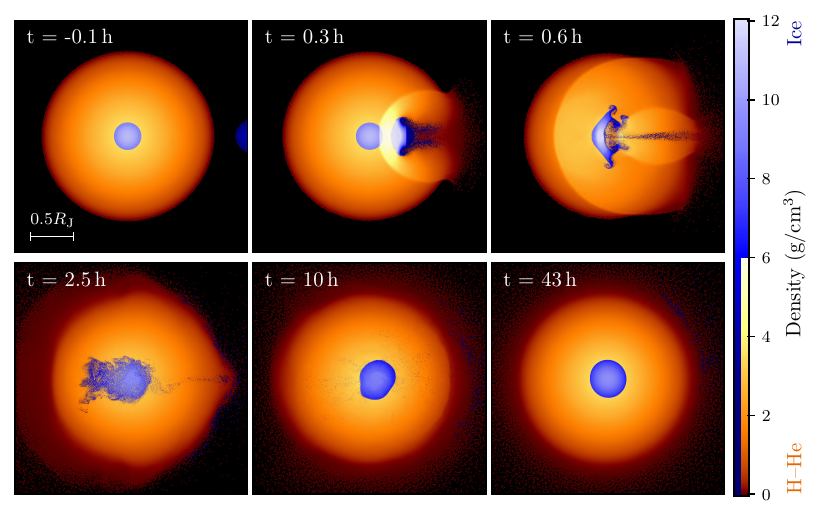}%
}\hfill
    \vspace{-3em}
	\caption{Snapshots from the fiducial, head-on impact on to Jupiter carried out using REMIX SPH. Individual SPH particles are plotted in cutaways from 3D simulations and are coloured by material-type and density. An animation of this impact is available at \url{www.icc.dur.ac.uk/giant_impacts/jupiter_remix_1e7.mp4} and a 3D-rendered animation of the equivalent $10^8$-particle resolution impact is available at \url{https://youtu.be/xkpZSNlrWTg}.}
	\label{Jupiter_fig:impact_remix}
    \vspace{-1em}
\end{figure*}

\section{Giant impacts}\label{Jupiter_sec:giant_impacts}

Here, we investigate dilute-core formation in simulations of: head-on impacts (\S\ref{Jupiter_subsec:fiducial}); isolated planets with a pre-constructed dilute core to test whether our simulation methods would in principle be able to produce a dilute core that is stable for the duration of our impact simulations (\S\ref{Jupiter_subsec:predilutecore}); impacts at a range of speeds and angles (\S\ref{Jupiter_subsec:speed_angle}); and impact simulations with initial conditions set up to closely replicate those of \citetalias{liu2019formation}, including an alternative version with more sophisticated EoS (\S\ref{Jupiter_subsec:L19_impact}).

\subsection{Fiducial scenario}\label{Jupiter_subsec:fiducial}

As a basis for investigations of impact configuration and numerical resolution in later sections, we consider a fiducial scenario of the head-on impact between a $10~M_\oplus$ impactor and a $308~M_\oplus$ proto-Jupiter with a $10~M_\oplus$ core, at the mutual escape speed of the two bodies. We choose to focus primarily on impacts with a two-layer proto-Jupiter with only layers of ice and H--He  and a single-layer, ice impactor. We do this to deliberately reduce both the number of density discontinuities and the size of the core--envelope density contrasts in the initial conditions, to remove barriers to forming a dilute core in our simulations. We use the more advanced CD21 H--He and AQUA EoS for these simulations.

\begin{figure*}
	\centering
{\includegraphics[width=\textwidth, trim={1mm 0mm 2mm 0mm}, clip]{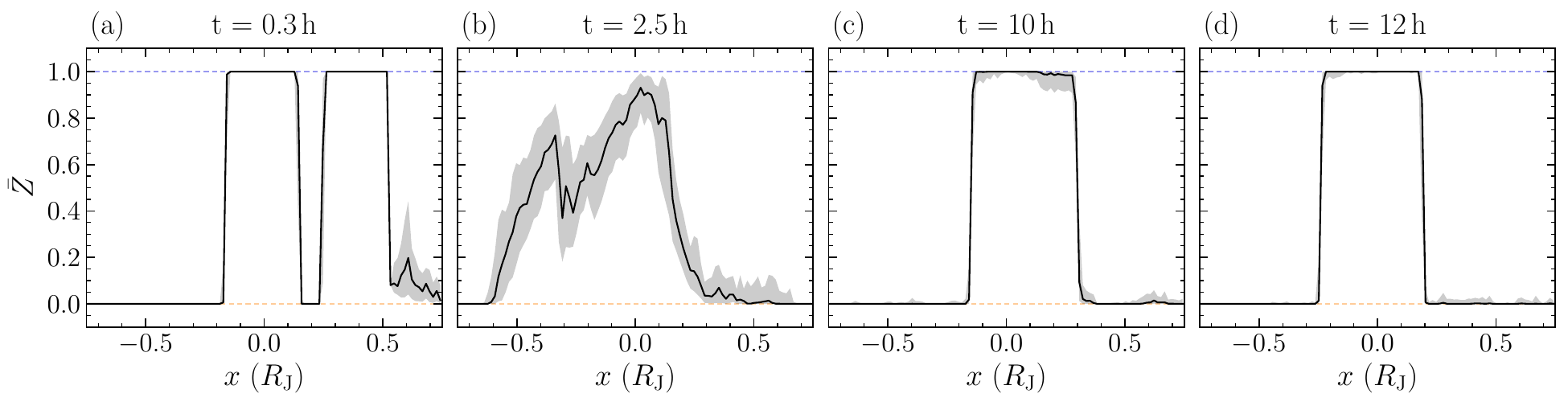}%
}\hfill	
    \vspace{-2em}
	\caption{Profiles of localised heavy-element mass fraction, $\bar{Z}$, sampled in a thin, 0.05~$R_{\rm J}$ radius cylinder along the axis of head-on impact, $x$, from the fiducial REMIX simulation. Panels correspond to times: (a) immediately prior to core-disruption by the impactor; (b) when heavy elements are mixed with envelope material; (c) when the core has settled to form a discrete boundary, although still mixed with some envelope material; (d) when core and envelope material have largely separated. Black, solid lines show the median particle value in 100 bins along the plotted region. Grey shading spans percentiles such that the enclosed regions correspond to 68\% of particles in each bin. The upper blue and lower orange dashed lines show heavy-element fractions of pure ice and hydrogen--helium, respectively. The $x$-axis is centred at the centre of mass of the system; deviations of the core position from $x=0$ are due to post-impact oscillations.}
	\label{Jupiter_fig:heavyelement_evolution}
\end{figure*}

We simulate the fiducial scenario using both REMIX and tSPH. In the REMIX simulation, ice particles can mix freely into the envelope, as seen in  Fig.~\ref{Jupiter_fig:impact_remix}. The core reaches a temporarily somewhat-mixed state, however, heavy elements rapidly settle under gravity to re-form a differentiated core over short time-scales of ${\sim}10$~h. The snapshots at 43~h show a later time where post-impact bulk-material oscillations have dissipated. No dilute core is produced, even with the improved treatment of mixing in the REMIX scheme.  
In the tSPH simulation, spurious surface tension-like effects are strong, suppressing mixing of ice and H--He particles. Heavy elements remain in a largely cohesive mass throughout the simulation, which settles to form a core with a discrete interface between the two different materials. A figure showing snapshots from the tSPH simulation is presented in Appendix \ref{Jupiter_app:numerical}.

We additionally perform REMIX simulations of this impact with resolutions $N$ = $10^5$--$10^{8}$ SPH particles in logarithmic steps of $10^{0.5}$. Although higher resolution allows turbulence to be resolved at lower length-scales, therefore extending the time materials take to separate, all of these simulations produce an undiluted core over the short time-scales simulated, as shown in Appendix \ref{Jupiter_app:numerical}.

Although the constituent equations of both REMIX and tSPH conserve energy, the kick-drift-kick algorithm with individual particle time-step sizes introduces slight variations in the system's total energy during the simulation. For tSPH and REMIX, these variations remain within 0.056\% and 0.073\% of the initial total energy, respectively.

\begin{figure}
	\centering
 \vspace{1.5em} 
{\includegraphics[width=0.48\textwidth, trim={2.5mm 0mm 2.5mm 0mm}, clip]{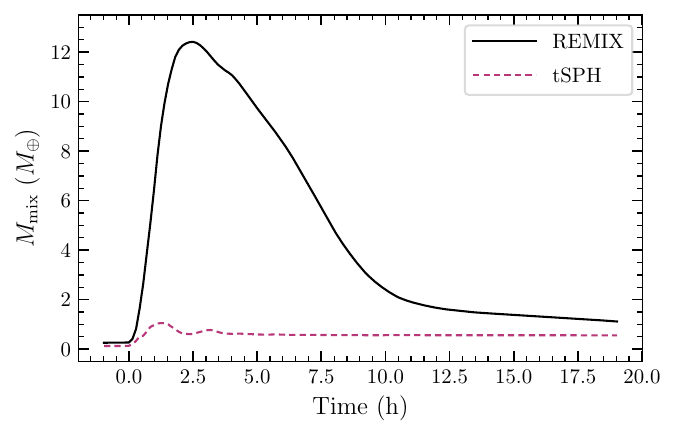}
}\hfill
    \vspace{-2em}
	\caption{Evolution of the total mass of mixed material, $M_{\rm mix}$, in simulations of head-on impacts on to Jupiter. The black, solid line shows results from a REMIX simulation and the pink, dashed line from a tSPH simulation. The total mass of heavy elements is  $20~M_\oplus$ in these simulations.}
	\label{Jupiter_fig:M_mix_evolution}
\end{figure}

The evolution of the local heavy-element fraction (Eqn.~\ref{Jupiter_eq:Z}) of the REMIX impact is shown in Fig.~\ref{Jupiter_fig:heavyelement_evolution}, for a thin cylinder aligned along the direction of the impact. At early times, $\bar{Z}$ reaches intermediate values as material mixes due to the erosion of the impactor as it travels through the envelope, as seen in Fig.~\ref{Jupiter_fig:heavyelement_evolution}(a), and due to the disruption of the core by the impact, the immediate aftermath of which is shown in Fig.~\ref{Jupiter_fig:heavyelement_evolution}(b). At 10~h, the core, not positioned at the centre of mass of the planet due to the oscillations, consists largely of heavy elements, and the core--envelope interface is already sharp. By 12~h the core is close to consisting purely of heavy elements. 

The evolution of the total mixed mass (Eqn.~\ref{Jupiter_eq:M_mix}) in these simulations is shown in Fig.~\ref{Jupiter_fig:M_mix_evolution}. There is considerably more mixing with REMIX than with tSPH. In the REMIX simulation, mixing peaks at a time 2.5~h after impact, at which time the core- and impactor-material particles have been maximally disrupted and mixed with the H--He envelope. After this time, the mass of mixed material falls as the system settles under gravity and materials separate, with material being largely separated by ${\sim}10$~h. We carry out simulations until later times to allow the large, dynamical oscillations to dissipate, although for the majority of this time the boundary of the core is already sharp and oscillations only affect its shape.

\subsection{Stability of a pre-constructed dilute core}\label{Jupiter_subsec:predilutecore}

We now address the possibility that material separation in the REMIX impact arises from numerical errors that would prevent these hydrodynamic methods from sustaining a non-transient dilute core over these time-scales no matter the scenario. To test this, we construct a planet with initial profiles of heavy-element fraction and density that match the post-impact planet produced in the simulation of \citetalias{liu2019formation} that produced a dilute core (from their Fig.~2a and the initial frame of their supplementary information video~5). These simulations are performed to assess whether a dilute-core structure can persist for the runtime of our impact simulations, rather than to infer the stability of the specific dilute core produced by the giant impact of \citetalias{liu2019formation}. Therefore, some evolution in the planet's radial profiles is not a concern, and is expected due to differences between the simulations, such as the number of materials and the treatment of mixed materials in the simulation methods.

The initial planet is constructed by placing ${\sim}10^7$ particles in a configuration that corresponds to the desired density profile. Each particle's material is set probabilistically based on the heavy-element fraction profile. For instance, at a radius where the heavy-element fraction is $Z=0.3$, a particle will have a 30\% chance of being assigned the AQUA ice EoS and a 70\% chance of the CD21 H--He EoS. Specific internal energies are then chosen such that their pressures satisfy hydrostatic equilibrium, with a pressure of $10^7$~Pa at the vacuum boundary, as this was found to give a relatively stable vacuum interface. Note that here, in keeping with the giant impact simulations of the previous section, we represent heavy elements only by ice, rather than by ice and rock, as done in the simulations of \citetalias{liu2019formation}. Therefore, the materials and thermodynamics of our initial planet are not directly equivalent to \citetalias{liu2019formation}'s post-impact planet; we are primarily focused on the comparative stability of a dilute-core structure rather than a specific planetary profile. Although these simulations are of a planet in isolation here, unlike the settling simulations we perform prior to impacts, we do not fix particle entropies. This approach more directly addresses whether a non-transient dilute core could, in principle, be produced in our impact simulations.

\begin{figure}
	\centering
 \vspace{1.5em} 
{\includegraphics[width=0.48\textwidth, trim={2.5mm 0mm 2.5mm 0mm}, clip]{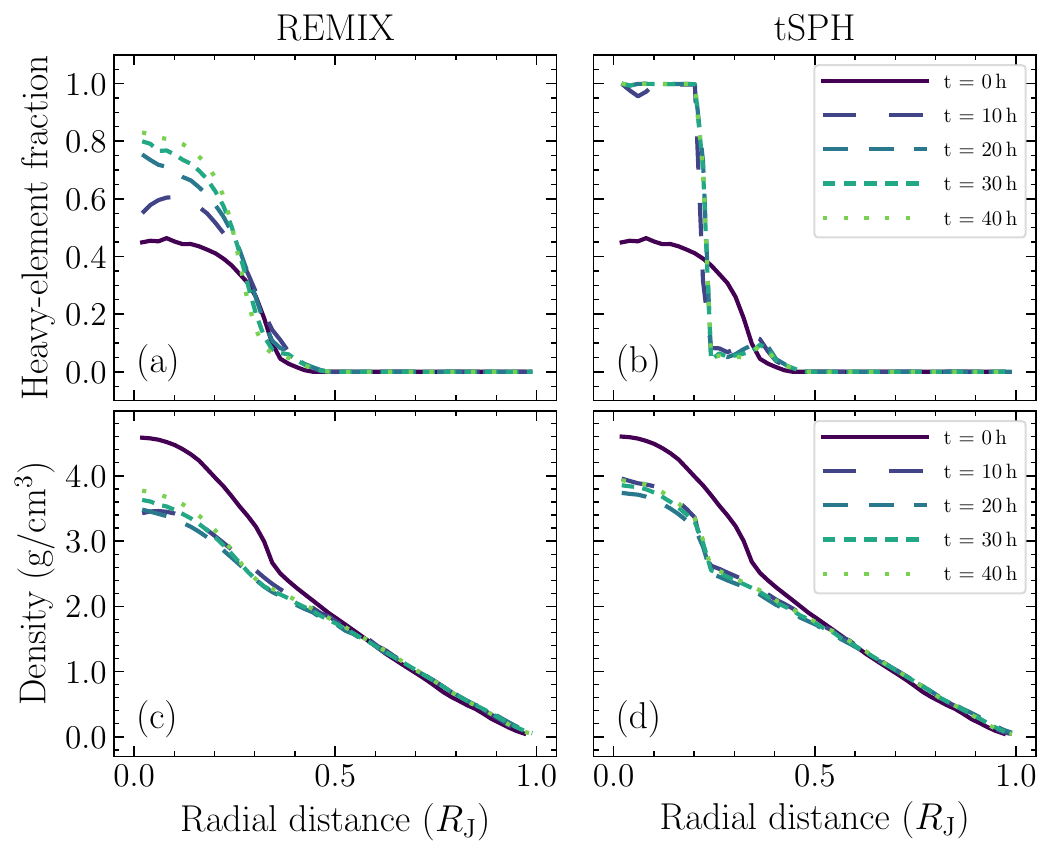}
}\hfill
    \vspace{-2em}
	\caption{Evolution of radial profiles of heavy-element fraction (a, b) and density (c, d) in simulations of a Jupiter-like planet with a pre-constructed dilute core. Simulations are carried out with REMIX (a, c) and tSPH (b, d) and profiles are plotted at five times to show the stability of a dilute core over the time-scales of our giant impact simulations. The plotted profiles show the mean values of the quantities within 50 radial shells.}
	\label{Jupiter_fig:predilute}
\end{figure}

The radial profiles of density and heavy-element fraction at five points in time in these simulations, carried out using REMIX and tSPH, are shown in Fig.~\ref{Jupiter_fig:predilute}. The times plotted are chosen to show the stability of a dilute core for the time-scale of the duration of our planetary impact simulations, as for an impact, the core rapidly settles to a differentiated state already by ${\sim}$10~h. With REMIX the core remains dilute with a smooth interface. Although the profiles themselves evolve in time in both cases, the profiles evolve less substantially between $t=20$--$40$~h than in the first 20~h. Further advancements in simulation methodology -- particularly in capturing mixing processes below the particle scale -- might improve the stability of the compositional gradients. At $t=40$~h, the planet has a central heavy-element fraction of $Z=0.83$ and the dilute-core structure extends out to $0.4$--$0.5~R_{\rm J}$. With tSPH, materials separate within the first 10~h to form an undiluted core, as in the impact simulations using either method.

\begin{figure}
	\centering
{\includegraphics[width=0.45\textwidth, trim={2.5mm 0mm 2.5mm 0mm}, clip]{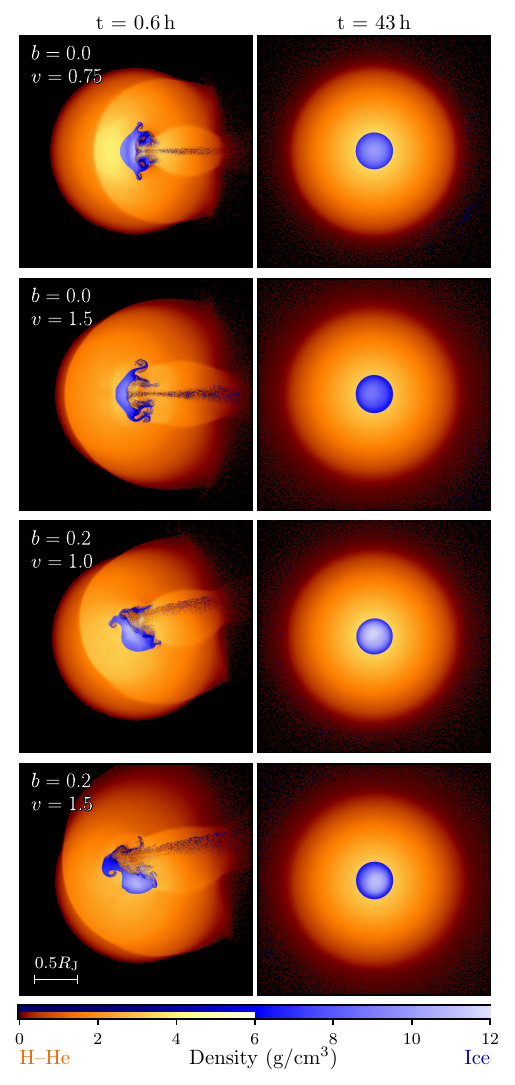}
}\hfill
    \vspace{-1em}
	\caption{Snapshots from REMIX simulations of giant impacts on to Jupiter for different impact parameters, $b$, and speeds, $v$. Impact velocities are scaled to the mutual escape speed of the two bodies. Times correspond to core-disruption and long after impact, when heavy elements have settled to form an undiluted core. Particles are coloured by their material-type and density.}
	\label{Jupiter_fig:speeds_angles}
\end{figure}

\subsection{Impact speed and angle}\label{Jupiter_subsec:speed_angle}

Although no stable dilute core is produced in the head-on scenario of \S\ref{Jupiter_subsec:fiducial}, this result could perhaps be sensitive to the speed and angle of the impact. The core of the planet is more likely to be disrupted in head-on or low-angle impacts, but impactor material might mix into the envelope more effectively by erosion in higher-angle impacts. Additionally, one might speculate that higher impact speeds may act to increase the initial material mixing. Or, conversely, perhaps lower speeds could lead to post-impact heavy-element distributions and internal energy profiles that are more stable to convection that may otherwise facilitate demixing of materials. Therefore, investigating a wide range of impact speeds and angles will enable us to examine the sensitivity of dilute core production to the impact configuration. 

For this parameter study, we use the same initial planetary bodies as in the fiducial scenario, although we run simulations with all additional combinations of four impact parameters and three impact speeds, listed in \S\ref{Jupiter_subsec:ICs}. The choices of $v = 0.75$~$v_{\rm esc}$ and $v = 1.5$~$v_{\rm esc}$ represent extreme scenarios to probe the sensitivity of our results to large changes in the impact kinematics. All of these simulations were performed using REMIX with $10^7$ particles.

Snapshots from impact simulations with different speeds and angles are shown in Fig.~\ref{Jupiter_fig:speeds_angles}. Although these examples constitute a small selection of the impacts simulated, they specifically correspond to speeds and angles in which the core is significantly disrupted. During the impacts, heavy elements mix into the envelope through both the erosion of the impactor and the disruption of the core, in particular for low impact angles. In head-on impacts energy is more effectively transferred to the core and so the post-impact core-density in these impacts is lower than for off-axis impacts. However, for all impact configurations, heavy elements settle over short time-scales to form an undiluted core with a discrete boundary to the H--He envelope.

\subsection{Direct Liu et al. (2019) comparison}\label{Jupiter_subsec:L19_impact}

\begin{figure*}
	\centering
\includegraphics[width=\linewidth]{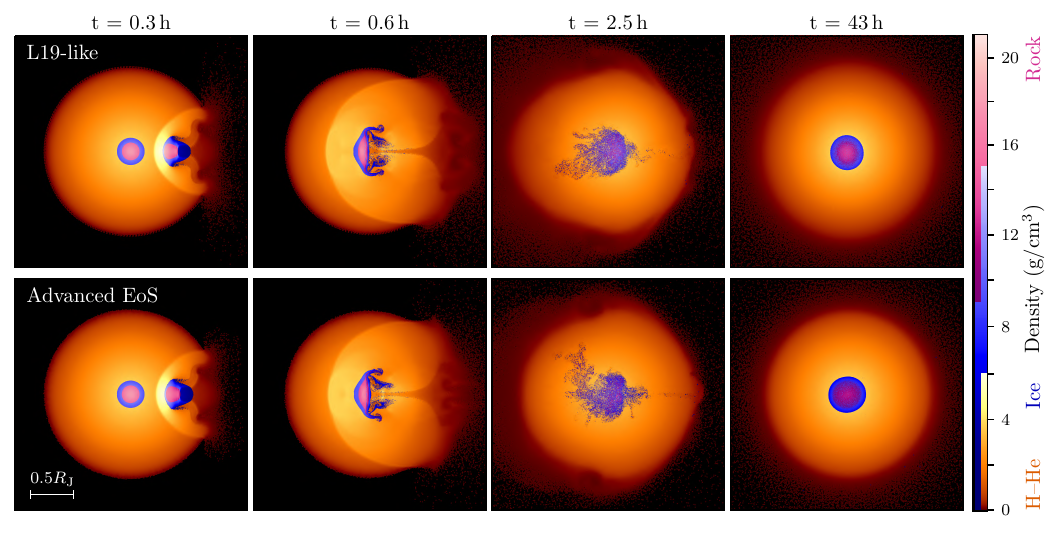}%
\vspace{-2em}
\caption{Snapshots from REMIX simulations of head-on impacts on to Jupiter set up to most closely follow the initial conditions of the impact of \citetalias{liu2019formation} that produced a dilute core (top) and the equivalent scenario with more advanced EoS (bottom). Particles are coloured by their material-type and density.}
\label{Jupiter_fig:3layer}
\end{figure*}

Here we simulate impacts that are set up to directly follow the impact of \citetalias{liu2019formation}, which they found to produce a well-mixed dilute core. By closely matching their initial conditions and the EoS used, any remaining differences in simulation outcomes should be primarily due to the numerical methods used in the simulations. We additionally simulate an equivalent impact using more advanced EoS for both core and envelope materials. This will allow us to investigate the sensitivity of our results to the EoS used.

In the simulations of \citetalias{liu2019formation}, planetary profiles are initially constructed using the SESAME EoS \citep{lyon1978sesame} and then swapped for Tillotson and ideal gas EoS for the impact simulations, replacing the initial particle internal energies to recover the SESAME pressure and density profiles (S.-F. Liu personal communication, 2020). For our subset of comparison simulations set up to match those of \citetalias{liu2019formation}, we therefore carry out a similar process by swapping the EoS from profiles calculated using ANEOS forsterite, AQUA, and CD21 H--He to Tillotson EoS and ideal gas with $\gamma = 2$. We verify that our proto-Jupiter profile closely matches the one used in \citetalias{liu2019formation}'s simulations, so the difference between using these EoS or SESAME to construct the pre-swapped profiles is minor. We find that planets with a swapped-in ideal gas envelope are not stable for the duration of settling simulations. Therefore, impact simulations that use these EoS are run without prior settling simulations. As in previous simulations, the proto-Jupiter has a total mass of  $308~M_\oplus$ and the impactor has a mass of $10~M_\oplus$. 

Snapshots from impacts of both the direct \citetalias{liu2019formation} comparison and the equivalent simulation with improved, more sophisticated EoS are shown in Fig.~\ref{Jupiter_fig:3layer}. Although there are small differences in dynamics during the course of these impacts, they each follow a similar evolution, both to each other and to all previous impacts simulated here. The core is disrupted and material temporarily mixes into the envelope, but heavy elements settle to form a discrete core--envelope interface over short time-scales. Some rock and ice core materials remain mixed with each other at later times, however they are not diluted by envelope material.

\section{Discussion}\label{Jupiter_sec:discussion}

\begin{figure*}
	\centering
{\includegraphics[width=\textwidth, trim={2.5mm 0mm 2.5mm 0mm}, clip]{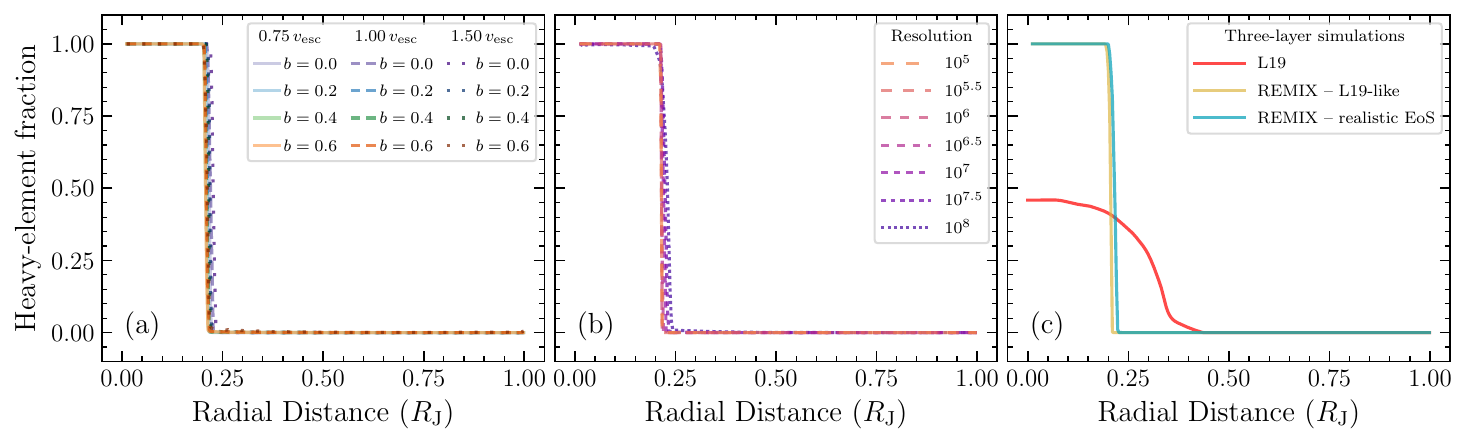}%
}\hfill
    \vspace{-2em}
	\caption{Radial profiles of heavy-element mass fraction at 43~h after impact, from simulations (a) at different impact speeds and angles; (b) with different numerical resolution; (c) set up to closely follow the initial conditions of \citetalias{liu2019formation}, including the equivalent profile from their simulation that produced a dilute core. Heavy-element fraction is measured by the ratio of heavy-element SPH particle mass to total mass in 300 radial shells.}
	\label{Jupiter_fig:heavyelements}
\end{figure*}

None of the giant impact simulations presented here produced a dilute core. The heavy-element mass fraction profiles of the post-impact planets from our simulations are plotted in Fig.~\ref{Jupiter_fig:heavyelements}. No dilute core was produced in our simulations (1) spanning a range of impact speeds and angles; (2) at different resolutions; and (3) between planets with different compositions. We carried out simulations both under conditions set up to directly mirror those of \citetalias{liu2019formation} and in conditions set up to facilitate mixing and remove potential barriers to it, in an attempt to offer the best chance of dilute core production. The red line in Fig.~\ref{Jupiter_fig:heavyelements}(c) shows the profile of post-impact heavy-element fraction from the simulation of \citetalias{liu2019formation} that produced a well-mixed dilute core with a central heavy-element fraction of $Z<0.5$. The categorical difference in our results and the simulation of \citetalias{liu2019formation} are likely due to differences in the simulation methodologies used.

Our primary suite of simulations were performed using REMIX SPH, an advanced SPH formulation that was developed specifically to improve the treatment of material mixing in SPH simulations. As a Lagrangian method where interpolation points move with the fluid velocity, we do not encounter the issues in regions of large bulk motion that are known to affect grid-based codes (e.g. Figs.~33 and 36 of \citealt{springel2010pur}). In this aspect, the methods used by \citetalias{liu2019formation} face potential limitations in the application of studying mixing in regions where there is large advection through the stationary grid points. The well-established overmixing in grid-based codes in regions of large bulk motion through the grid may be an explanation for their results: in their simulations, the core rapidly mixes into the envelope as it is accelerated by the impactor. 

Although we demonstrate significant improvements in the treatment of mixing in our simulations compared with simulations carried out with a tSPH scheme, we have not considered mixing of material below the length-scale of SPH particles \citep{greif2009chemical} or the chemical reactions that might affect the evolution of materials as they mix. The extension of the simulation methods to include these potentially important mechanisms may also help address the resolution dependence of mixing observed in these highly turbulent scenarios (Appendix \ref{Jupiter_app:numerical}). Further progress in quantifying the discrepancy between simulated scenarios and their physical analogues -- which remains challenging due to the chaotic nature of impact dynamics and the lack of observational constraints in this regime -- would help to better assess the sensitivity of large-scale simulation outcomes to both physical processes and numerical uncertainties. Future work should focus on developing improved numerical methods to model mixing and demixing processes occurring below the resolution scale of SPH particles. Different simulation approaches should be compared in isolated test scenarios, like those we present in \S\ref{Jupiter_sec:instabilities}, to further analyse the demixing processes observed during the simulated impacts, providing a physically motivated framework for quantifying the differences in the material separation mechanisms with different numerical methods.

However, as shown in \S\ref{Jupiter_subsec:predilutecore}, a dilute core structure can be sustained in our REMIX simulations with limited material separation over the time-scale of the impact simulations. This suggests that our impact simulations -- despite lacking sub-particle scale mixing and chemistry -- could in principle produce a dilute core if a more stable configuration were to be reached. Therefore, the absence of a dilute core in our impact simulations appears due to the giant impacts' inability to disrupt the core to a more stable diluted state.

In addition to the hydrodynamic methods, there are differences in the approaches taken in the calculation of self-gravity. In our simulations, we employ the fast multipole method \citep{greengard1987fast, cheng1999fast}, which partitions the simulation domain into a hierarchical tree of spatial cells. Gravitational interactions between nearby particles are calculated directly, while interactions over larger distances are approximated by multipole expansions of particle groups within cells \citep{schaller2024swift}. In contrast, \citetalias{liu2019formation} estimate the full gravitational potential using a single multipole expansion centred on the system’s centre of mass \citep{liu2015giant}. Choosing the centre of mass as the expansion centre, rather than alternatives like the location of peak density \citep{sellwood1987art} or a ``square-density-weighted mean location'' \citep{couch2013improved}, can introduce errors in gravitational force estimates, particularly when the centre of mass deviates significantly from the peak-density location. During the giant impact simulations by \citetalias{liu2019formation}, the disruption of the core of heavy elements coincides with a significant shift of this high-density region from the centre of mass, which is primarily set by the much larger mass of the envelope. Therefore it is not clear whether error in the calculation of gravity may also play a role in the rapid mixing of material during core disruption, observed in their simulation.

Separately from this discussion of the numerical methods, it should be noted again how extreme and specific the dilute core-producing impact simulation of \citetalias{liu2019formation} is: the impact conditions require the head-on impact of a $10~M_\oplus$ on to an almost fully formed Jupiter that has accreted almost all of its final envelope mass yet only half its heavy elements. This, combined with the inference of a dilute core in Saturn \citep{mankovich2021diffuse} in addition to Jupiter, might suggest that it is more likely that dilute cores are produced as part of the extended processes that underlay the formation and evolution of giant planets, rather than through low-likelihood stochastic events \citep{helled2024fuzzy}. 

Recent models of giant planet formation indicate that composition gradients naturally arise during the formation process \citep{helled2017fuzziness, lozovsky2017jupiter, Stevenson+2022} and that an extended period of planetesimal accretion could deliver sufficient energy to delay runaway gas accretion \citep{venturini2020jupiter}. Since the material delivered by runaway gas accretion is less rich in heavy elements, offsetting this phase could allow the compositional gradients that constitute the dilute core to extend further from the planet's centre. Alternatively, under certain conditions, thermal convection after giant planets have formed could lead to the convective mixing of core and envelope material \citep{moll2017double}. These formation pathways are perhaps more promising than a single, low-probability giant impact, which our results suggest could be unable to produce a dilute core even under the extreme impact conditions considered here.

\section{Conclusions}\label{Jupiter_sec:conclusions}

We have presented results from REMIX SPH simulations of giant impacts on to Jupiter to investigate the feasibility of this as the process by which the planet’s dilute core was formed. We varied impact speed, angle, numerical resolution, the number of layers in the pre-impact planets, and the EoS used to represent proto-Jupiter and impactor materials. The impact dynamics in all simulations followed the same trend: initial disruption and partial mixing, followed by settling and re-formation of an undiluted heavy-element core on ${\sim}10$~h time-scales. However, our simulations do not account for mixing below the resolution scale of SPH particles or the effects of chemical reactions, which could, in principle, influence the large-scale outcomes of the simulations. The first of these may also be required for numerical convergence of these highly turbulent impact scenarios.

Our results contrast with the simulation of \citet{liu2019formation} that produced a highly dilute core with a central heavy-element fraction of $Z<0.5$ and a smooth transition to the envelope. Their result is potentially an artefact of numerical issues, such as the well-established overmixing in grid-based codes in regions of large bulk motion through the grid.

The REMIX SPH scheme was specifically designed to improve the treatment of mixing and instability growth. Despite our approach offering favourable conditions and spanning a wide parameter space, dilute cores were not produced in any of our simulations. This result, reinforced by observations that suggest that dilute cores are not unique to Jupiter, offers no support for the hypothesis that a single, extreme giant impact is the origin of dilute cores in giant planets.

\section*{Acknowledgements}

TDS acknowledges support from STFC grants ST/T506047/1 and ST/V506643/1.
VRE\ and RJM\ are supported by Science and Technology Facilities Council (STFC) grant ST/X001075/1.
JAK\ acknowledges support from a NASA Postdoctoral Program Fellowship administered by Oak Ridge Associated Universities.
The research in this paper made use of the \swift open-source simulation code \citep{schaller2024swift}.
This work used the DiRAC@Durham facility managed by the Institute for Computational Cosmology on behalf of the STFC DiRAC HPC Facility (www.dirac.ac.uk). The equipment was funded by BEIS capital funding via STFC capital grants ST/K00042X/1, ST/P002293/1, ST/R002371/1 and ST/S002502/1, Durham University and STFC operations grant ST/R000832/1. DiRAC is part of the National e-Infrastructure.

\section*{Data Availability}

The data underlying this article will be shared on reasonable request to the corresponding author.

\renewcommand{\thefigure}{A\arabic{figure}}
\setcounter{figure}{0}  

\begin{figure*}[h]
	\centering
\includegraphics[width=\linewidth]{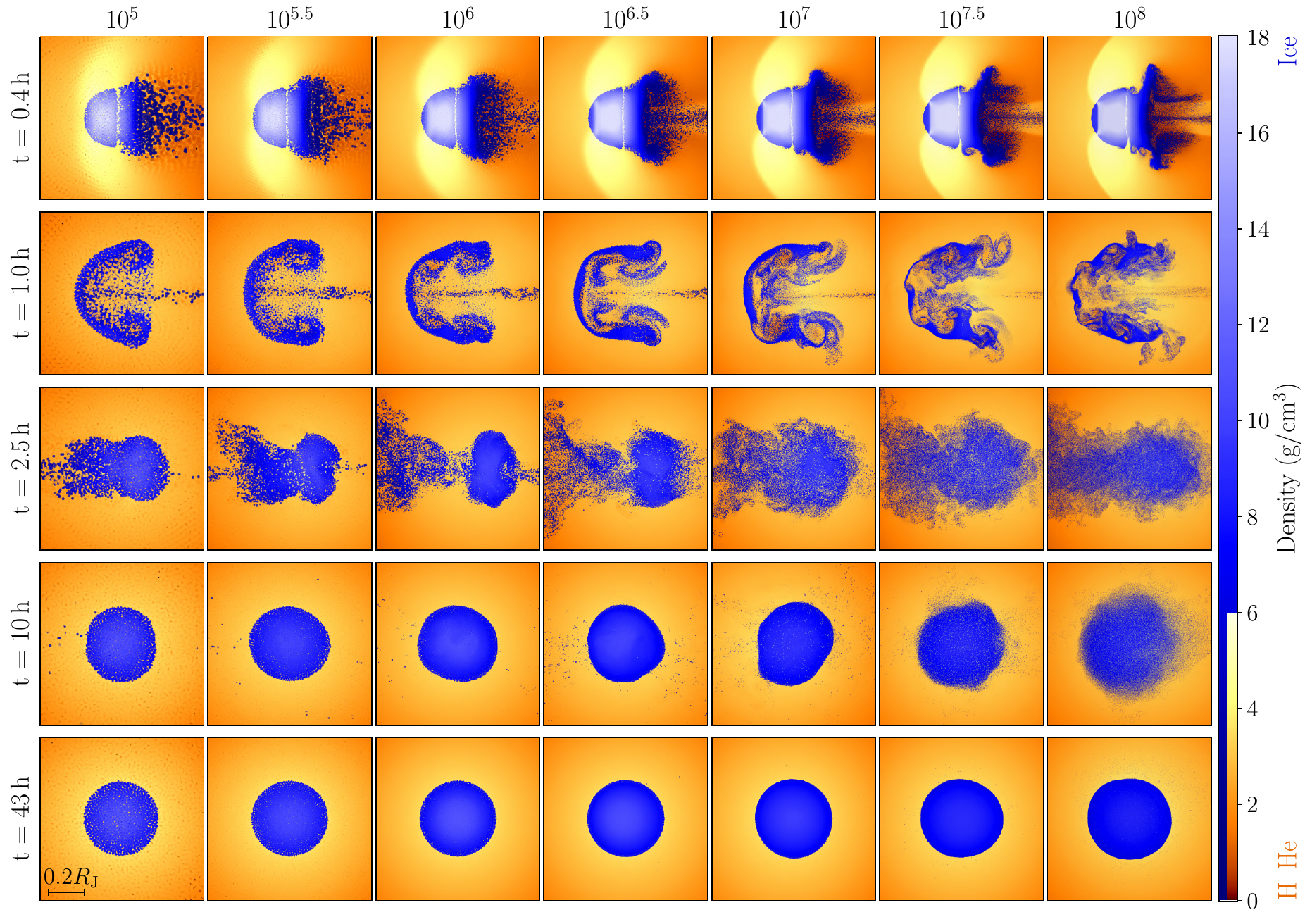}%
\vspace{-1.5em}
\caption{Snapshots from REMIX simulations of head-on impacts on to Jupiter, carried out with different numerical resolutions between $10^{5}$ and $10^{8}$ SPH particles. The five times plotted illustrate material mixing at different points during core-disruption and after the heavy elements have settled to form an undiluted core. The plotted region is centred on the centre of mass of the heavy elements. Particles are coloured by their material-type and density.}
\label{Jupiter_fig:resolutions}
\end{figure*}

\begin{figure*}
	\centering
{\includegraphics[width=\textwidth, trim={2.5mm 0mm 2.5mm 0mm}, clip]{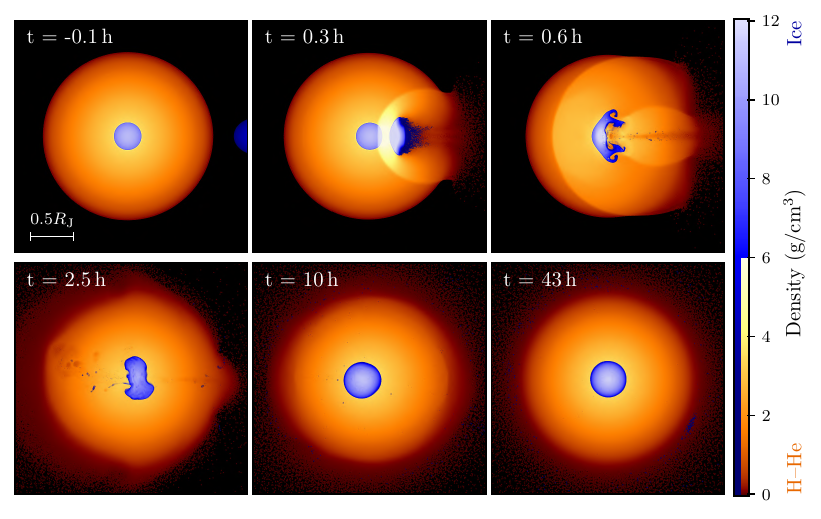}%
}\hfill
    \vspace{-2em}
	\caption{Snapshots from the fiducial, head-on impact on to Jupiter carried out using a tSPH formulation, shown as in Fig.~\ref{Jupiter_fig:impact_remix}. Individual SPH particles are plotted in cutaways from 3D simulations and are coloured by material-type and density. An animation of this impact is available at \url{www.icc.dur.ac.uk/giant_impacts/jupiter_tsph_1e7.mp4}.}
	\label{Jupiter_fig:impact_trad}
\end{figure*}



\bibliographystyle{mnras}
\bibliography{bibliography} 




\appendix

\section{Numerical aspects of impact simulations}\label{Jupiter_app:numerical}

Here, we consider some effects of the numerical aspects of the fiducial giant impact simulation presented in \S\ref{Jupiter_subsec:fiducial}. We briefly investigate how the numerical resolution influences the outcome of the simulations. Additionally, we present results from a simulation that was performed with tSPH to allow for a direct comparison between the treatment of mixing in simulations performed with different SPH formulations.

Numerical resolution determines not only the minimum length-scale probed in simulations, but can also significantly influence the accuracy of simulated fluid behaviour at all length-scales. 
In particular, even large-scale outcomes of SPH simulations of giant impacts can remain unconverged at standard resolutions of $10^{5}$--$10^{6}$ particles \citep{Genda+2015,Hosono+2017,kegerreis2019planetary,kegerreis2022immediate}.
Not only does the computational efficiency of the \swift code allow us to simulate giant impacts at far higher resolutions, but the REMIX scheme has been demonstrated to improve numerical accuracy such that convergence can be achieved at lower resolutions than in equivalent tSPH simulations \citep{sandnes2025remix}.

Here, we test whether we achieve numerical convergence in dilute core production -- or lack thereof -- in our head-on fiducial impact. We carry out simulations at resolutions of $10^n$ with $n = 5$--$8$, in steps of 0.5, SPH particles. All these simulations were performed using REMIX. The mixing of heavy elements into the H--He envelope in simulations of different resolutions is shown in Fig.~\ref{Jupiter_fig:resolutions}. We present five snapshots in time: two illustrating the initial disruption of the core by the impactor; a third capturing a moment of significant material mixing; a fourth at a time where heavy elements have largely re-settled, although more mixing is still present in the higher resolution simulations; and a fifth depicting the later stage when post-impact oscillations have dissipated and a distinct, undiluted core has been produced for all resolutions. The mixing of particles of different materials is observed for all resolutions. Increasing resolution allows the simulation to capture mixing and instability growth at smaller length-scales, with KHI growing at the shearing interface as predicted in \S\ref{Jupiter_subsec:kh}. Because of these turbulent effects, as resolution is increased, large-scale features become less symmetric about the impact axis and are more significantly disrupted by chaotic fluid behaviour at smaller scales. Since the turbulent mixing is captured at shorter length-scales with increased resolution, we find that the heavy elements take a slightly longer time to settle in the higher resolution simulations and we therefore do not achieve numerical convergence in the time-scale of demixing. Nevertheless, in all these simulations an undiluted and settled core is produced well within the short time-scales of these impact simulations.

In these simulations, we do not model mixing below the resolution scale of individual SPH particles, as each particle retains a fixed material type throughout the simulation. Achieving numerical convergence of the mixing and demixing processes in these particular highly turbulent impact scenarios might require modelling the evolution of particle composition with time, and subsequently accounting for the resulting changes to the particles' EoS.

We additionally perform a simulation of the fiducial scenario presented in \S\ref{Jupiter_subsec:fiducial} using the more tSPH formulation that was also used in the comparison fluid instability tests of \S\ref{Jupiter_sec:instabilities}, with the same resolution of $10^7$ particles. Snapshots from this impact are shown in Fig.~\ref{Jupiter_fig:impact_trad}. The panels of this figure are at the equivalent times to those in Fig.~\ref{Jupiter_fig:impact_remix}. The spurious surface tension-like effects at material boundaries here act to prevent mixing of the different materials material. Heavy elements remain largely unmixed throughout the course of the simulation and no dilute core is produced.


\bsp	
\label{lastpage}
\end{document}